\documentclass[fleqn,usenatbib]{mnras}

\usepackage{newtxtext,newtxmath}

\usepackage[T1]{fontenc}

\DeclareRobustCommand{\VAN}[3]{#2}
\let\VANthebibliography\thebibliography
\def\thebibliography{\DeclareRobustCommand{\VAN}[3]{##3}\VANthebibliography}

\usepackage{graphicx}	
\usepackage{amsmath}	

\title[CIRs over a stellar lifetime]{Corotating Interaction Regions (CIRs): evolution over a solar lifetime}

\author[Rose F. P. Waugh et al.]{
Rose F.P. Waugh,$^{1}$\thanks{E-mail: rw47@st-andrews.ac.uk (RFPW)}
Moira M. Jardine,$^{1}$\thanks{E-mail: mmj@st-andrews.ac.uk (MJ)}
\\
$^{1}$School of Physics and Astronomy, University of St Andrews, North Haugh, St Andrews, Fife, Scotland, KY16 9SS
}

\date{Accepted XXX. Received YYY; in original form ZZZ}

\pubyear{\the\year{}}

\begin{document}
\label{firstpage}
\pagerange{\pageref{firstpage}--\pageref{lastpage}}
\maketitle

\begin{abstract}
Corotating Interaction Regions (CIRs) are persistent structures in stellar winds that arise from the interaction between fast and slow wind streams. They are known to generate shocks and high-energy particles, potentially influencing the erosion of planetary atmospheres and space weather conditions. Although extensively studied for the present-day Sun, their evolution over a star’s lifetime and implications for planetary environments remain less explored. We model the evolution of CIRs around a solar-mass star using a rotational evolution framework and assess how their location, shape, and particle distributions vary with stellar age. We find that CIRs form closer to the star during early spin-up phases and migrate outward during spin down, with the minimum CIR radius inversely related to rotation rate. We show that the distribution of high-energy particles produced in CIR shocks varies significantly with age. During the Hadean period, when Earth's atmosphere evolved significantly, CIRs may have generated a number of energetic particles that is $10^3$ to $10^7$  times greater than for the present-day Sun. The frequency and strength of CIR-planet interactions also peak during early rapid rotation phases. Furthermore, we demonstrate from this preliminary study that, for stars with mass less than 1.4 $M_{\odot}$, whilst CIRs can form within the habitable zone, their shocks always form beyond it. This suggests that energetic particle impacts may rain inward from more distant regions, as with the present-day Sun. These findings have implications for habitability and the evolution of atmospheres since these particles can alter the chemistry and escape rate of atmospheres.
\end{abstract}




\section{Introduction}

\subsection{The importance of stellar winds}
Stellar winds govern the stellar evolution of low mass stars through the angular momentum that they remove. The spin down that this causes influences the stellar magnetic field through the stellar dynamo, which then feeds back into altering the stellar wind, as well as the space weather and environment of orbiting planets, thus the wind is important to both the star and host exoplanets~\citep{Zendejas2010,See2014,Vidotto2015,Strugarek2016,Reville2024,Airapetian2020}. Whilst the solar wind is well studied, winds on other low mass stars can not be directly observed due to their low densities, but their existence can be inferred indirectly~\citep{Wood2005,Wood2021}. The environment generated by the solar wind is known to cause geomagnetic storms of Earth (~\cite{Tsurutani1997,Watari2023,Liu2019} and references within) whilst for exoplanets, stellar winds may be able to strip the atmosphere away~\citep{Harbach2021,Holmstrom2008,Khodachenko2012,Kislyakova2014,Alvarado2016}. 
Planetary magnetospheres can be compressed by the stellar wind, and this compression and the shape of the compression are determined by the planetary magnetic pressure and the dynamic pressure of the wind relative to the escaping atmosphere, the interaction of which has been well studied~\citep{Schneiter2007,Caro2014,DaleyYates2017,McCann2019,Vidotto2020}. Winds, alongside flares, coronal mass ejections and high XUV radiation, are processes which could cause atmospheric loss and chemical changes in the atmospheres of exoplanets~\citep{Linsky2019,Airapetian2020}.

\begin{figure}
    \centering \includegraphics[width=1\columnwidth]{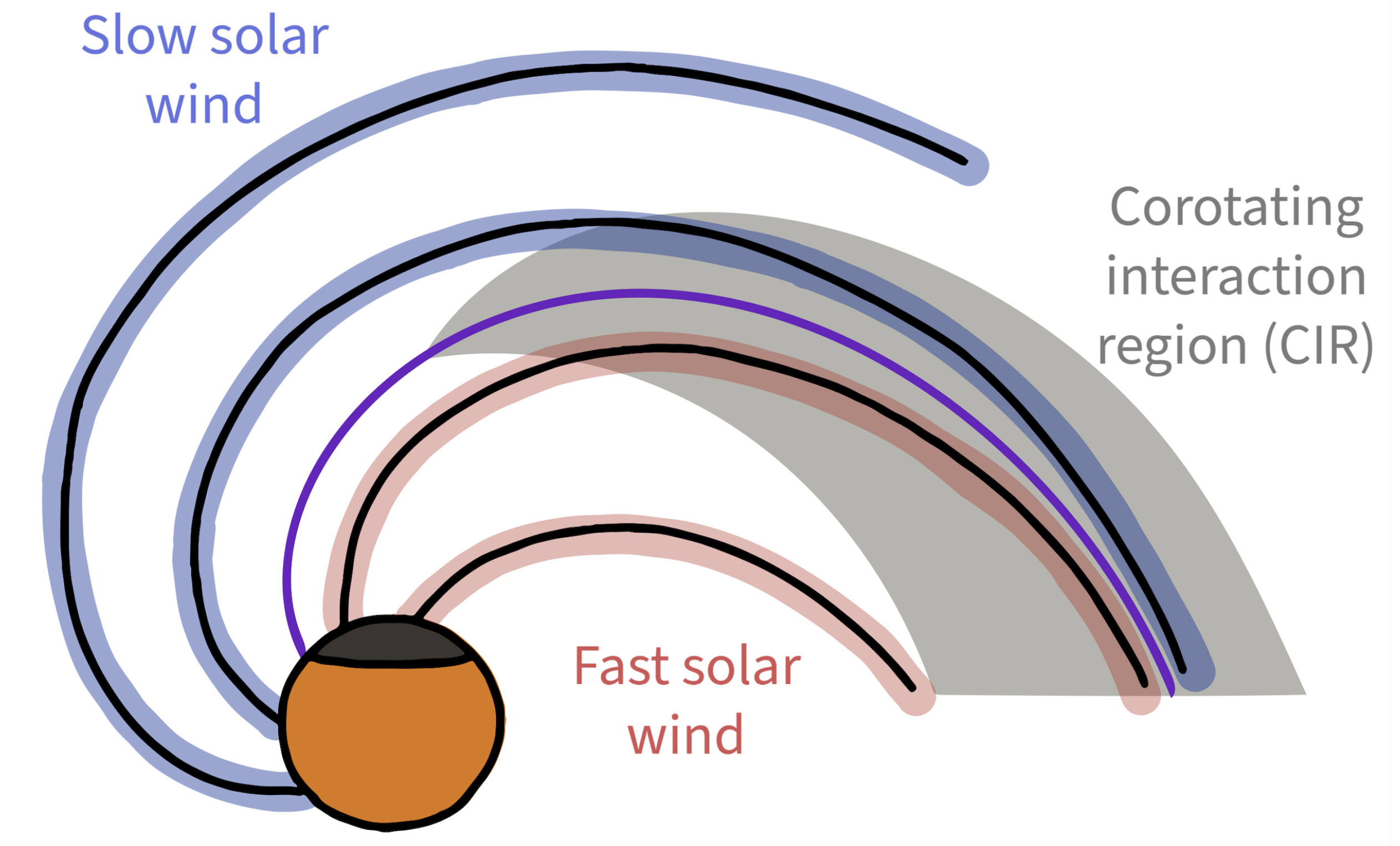}
    \caption{Schematic of a corotating interaction region (grey) forming at the boundary between the fast (orange) and slow (blue) winds.}
\label{fig:cartoon1}
\end{figure}

\subsection{CIRs}
Corotating interaction regions (CIRs)~\citep{Belcher1971,Smith1979} are regions of increased density and magnetic field strength that occur when fast and slow streams of the solar wind collide. A schematic is shown in Figure~\ref{fig:cartoon1}. They are often referred to as stream interaction regions (SIRs) and classified as CIRs when they are stable over multiple solar rotations. They form along the leading edge of the fast wind stream as a compression region, corotating with the Sun, and may last for several months. These compressed regions steepen into shocks when the fast wind is supersonic relative to the slow wind. Often two shocks are present; a \textit{forward shock} forming ahead of the CIR and a \textit{reverse shock} behind. These shocks typically form by about 2 AU within our Solar System~\citep{Gosling1976,Gazis1983}, although they have been observed to have formed below 1 AU~\citep{Richardson1984,Berdichevsky2000}.

\subsection{High Energy Particles and CIRs}
High energy particles are generated in the Sun by various mechanisms, such as from flares or coronal mass ejections (CMEs). Both are known to generate GeV particles, and the consequences to us in our Solar System include damage to satellites and health threats to astronauts. For planets orbiting active stars, CMEs and flares could set the conditions for atmospheric escape, whilst penetrating the atmosphere with energetic particles which may be important for biological processes~\citep{kay2019}. However, CIRs also generate high energy particles, albeit at lower energies (MeV rather than GeV) \citep{Borovsky2006}. Around the present day Sun, ions are accelerated to about 20 MeV/nucleon in the vicinity of CIRs~\citep{Richardson2004}. How the Sun compares to other low mass stars, and how this behaviour on the Sun has varied throughout its lifetime, is yet to be established and is considered within this paper in Section 3. Particles from the suprathermal tail of the solar wind are accelerated by the shocks of the CIR, but also in nonshock processes, such as by turbulence within the CIR~\citep{Richardson2004}.\\

The consequences of these high energy particles and high pressure CIRs on planets within our own Solar System can be significant. \cite{Dubinin2009} showed that the high pressure regions of CIRs within the wind can penetrate the magnetosphere of Mars, leading to erosion of the atmosphere. Whilst coronal mass ejections (CMEs) typically generate higher energy particles than CIRs, \cite{Edberg2010} showed that all pressure pulses lead to loss of ions from the Martian atmosphere, regardless of whether they stem from CIRs or ICMEs (interplanetary CMEs). It has also been suggested that CIRs may be more of a driver of Martian ionosphere loss than CMEs in the declining phase of the solar cycle~\citep{Ram2023}.

\subsection{Cosmic ray modulation}
Galactic cosmic rays (GCR) are energetic charged particles, stemming from objects such as supernovae remnants. Whilst some are modulated by the heliospheric magnetic field~\citep{Potgieter1998,Potgieter2013}, GCRs of energies less than 0.5 GeV reach the inner heliosphere~\citep{Lockwood1996}. Variations in GCRs that reach Earth have long been known to be associated with geomagnetic storms~\citep{Forbush1937} and the relationship between GCR temporal variations and corotating features of the solar wind has been studied extensively~\citep{Lockwood1960,Iucci1979,Duggal1981,Venkatesan1982,Burlaga1984,Newkirk1985,Richardson1996,Richardson1999,Dumbovic2011,Badruddin2016}. The \textit{Forbush decrease} occurs in periods when the solar wind sweeps away GCRs from the Earth, thus causing a dip in the GCR intensity that we experience by up to 20\% ~\citep{GRIEDER2001}. It has been suggested that cosmic ray modulation could be important in the development and continuation of life on Earth. Life has evolved on Earth in the presence of cosmic rays, but this radiation has not been constant over this time period~\citep{Karam2003,Dartnell2011,Melott2011}. It is not clear what exact role cosmic rays played in the development of life on Earth, although cosmic rays are known to cause damage to DNA~\citep{Atri2014} and influence exoplanetary atmospheres~\citep{RogersLee2020,RogersLee2021,Herbst2019,Herbst2024}. All CIRs in our Solar System depress the cosmic ray intensity, with the depression onset generally mapping to the stream leading edge and the maximum depression occurring at the CIR trailing edge~\citep{Richardson2004}. 
If the size, strength, or frequency of CIRs across the Sun's lifetime vary, then this could result in varying cosmic ray modulation over the Earth's lifetime. The modulation of cosmic rays may also differ across the stellar spectrum.

\subsection{The Hadean Eon}
The Hadean eon is the first of the eons in planet Earth's history, beginning with Earth's formation and lasting about 500 Myrs. During this time, the planet evolved from having a molten surface to a solid crust. The accompanied cooling allowed for the first oceans to condense and at the end of the Hadean period, the primary atmosphere of the Earth progressed to a secondary atmosphere~\citep{Zahnle2010}. Primary atmospheres are those that are originally captured from the solar nebula and thus are composed mainly of atomic hydrogen and simple hydrides such as water, methane and ammonia~\citep{Zahnle2010,Hayashi1979,Lewis1985,Ikoma2006}. These atmospheres are still present in the gas giants of our Solar System and likely were present for the inner planets. For Mars-like planets or smaller, this atmosphere can be easily lost by a single collision with another planetesimal, leading to the increase in thermal velocities of the atmosphere above the escape speed~\citep{Zahnle2006}. However, for Earth-sized planets, escape of the primary atmosphere is more difficult~\citep{sekiya1980,sekiya1981}.
Secondary atmospheres are more sophisticated, dominated by gases other than hydrogen~\citep{Hoolst2019}. The composition and mass of these atmospheres depend on the outgassing from the planetary interior, surface interactions, and the escape of this material into space~\citep{Hoolst2019}.\\
The dominant mechanisms for escape of the atmospheres has been shown in our Solar System to vary between the inner planets. On Venus, Earth and Mars, atmospheric escape is caused by the solar wind and XUV, as well as through evolution of the atmospheric composition itself (since some chemicals are more prone to escape than others)~\citep{Gronoff2020,Lammer2018}. On Earth, atmospheric loss occurs at the poles as an outflow along open field lines; Mars experiences atmospheric loss as an outflow of hot atoms, whilst on Venus it occurs via ion pickup~\citep{Lammer2018,Amerstorfer2017,Barabash2007}. This difference relates to the intrinsic magnetic field of Earth, which is not present on Venus and Mars. On these planets the atmospheric escape occurs at a boundary layer of the induced magnetosphere~\citep{Barabash2007,Gunell2018,Haaland2015}. The magnetosphere is important for preventing erosion by the solar wind, but its efficiency is dependent on the extent to which energy exchange within the atmosphere is prevented~\citep{Blackman2018}. The frequency and strength of CIRs may vary over the time period in which planetary atmospheres evolve, and this could particularly be the case when the primary atmosphere is present. The role of CIRs and any energetic particles they produce on the atmospheric loss is as yet unclear.\\

\subsection{CIRs with age}
Whilst CIRs are well studied within our Solar System, they have not yet been systematically considered in studies of other low-mass stars or indeed throughout the Sun's evolution. It is unclear if CIRs are more common at certain periods within a star's lifetime (and, therefore, planetary lifetime). It is also unclear how the distribution of particles produced within the shocks varies with stellar age. Could CIRs be a more significant aspect of space weather for younger stars than they are on the present day Sun? The work in this paper tackles these questions.



\section{How do CIR locations vary with stellar age?}
The minimum CIR location can be calculated from the equation below~\citep{Waugh2025}
\begin{equation}
    r_{CIR} = \Delta \phi\frac{u_{slow}}{\Omega},
    \label{eqn:rcir}
\end{equation}
where $u_{slow}$ is the slow wind velocity, $\Omega$ is the stellar rotation rate and $\Delta \phi$ is the angular separation of the sources of the fast and slow wind streams and so is a measure of the field complexity. This was discussed in detail in~\cite{Waugh2025} (hereafter Paper 1) and therefore will not be focussed on here.\\

\begin{figure}
    \centering \includegraphics[width=1\columnwidth]{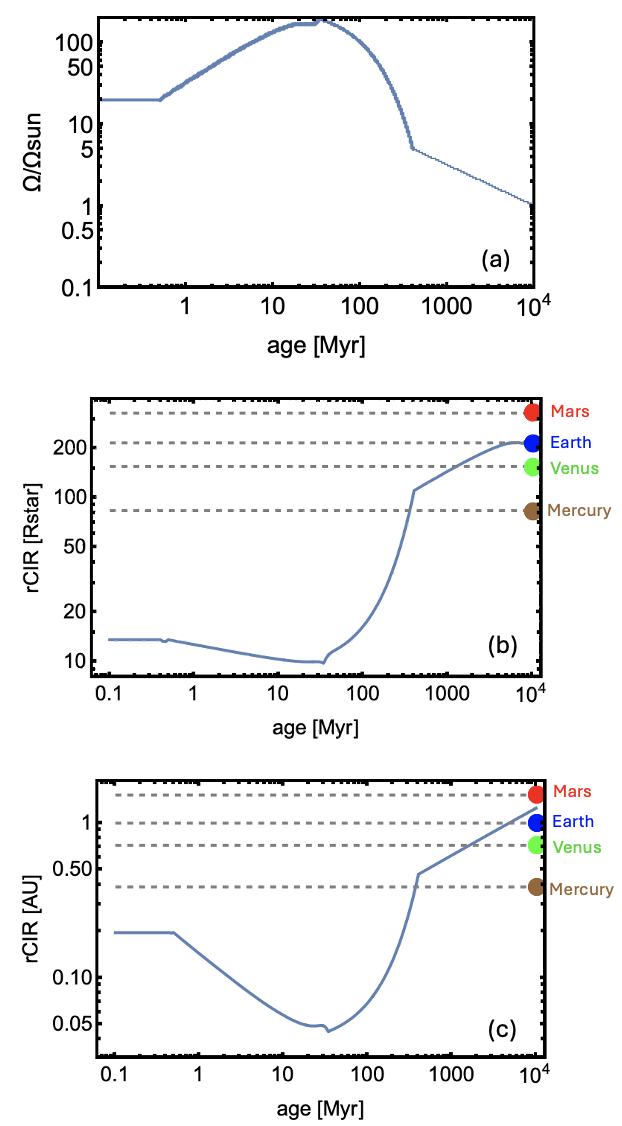}
    \caption{(a) Simple rotational evolution curve, generated from the Baraffe track for a solar mass star \citep{2015A&A...577A..42B}. (b) The minimum CIR radius supported with stellar age, with the orbits of Mercury and Venus shown by the dashed lines and brown and green points, respectively. (c) As in (b) but shown in units of AU, since the stellar radius also varies with age.}
\label{fig:cirwithage}
\end{figure}

The rotation rate is modelled here using a simple rotational evolution curve (as shown in Figure~\ref{fig:cirwithage}(a)). This curve is constructed from four functions, relating to; disc locking, stellar contraction and spin up, saturated spin down and Skumanich spin down. During disc locking, the star has a constant rotation rate, before spinning up in the second stage as the stellar core contracts. This second stage is modelled assuming angular momentum is conserved at each step: $\Omega_n r_n^2 = \Omega_{n+1} r_{n+1}^2$, where $\Omega$s and $r$s are the rotation rate and stellar radius, respectively. The initial step is before contraction begins, and the variation of the stellar radius with age was taken from \citet{2015A&A...577A..42B}.
In the third stage of saturated spin down, assuming a Weber-Davis model~\citep{weberdavis1967}, the rotation rate can be modelled by $d\Omega/dt \propto -\Omega$, whilst in the fourth regime the expression for Skumanich spin down is applied: $\Omega \propto t^{-1/2}$~\citep{skumanich1972}. Here we take a rapid rotator with an initial rotation rate of $20\Omega_{\odot}$, which corresponds to a star at the 90th percentile in rotation rate~\citep{Bouvier}.\\

We choose this simple, parameterised approach since it reveals the scalings in a transparent way that is easily reproducible by the reader. Our results demonstrate the range of behaviours that might be expected and could form a basis and guide to more sophisticated treatments of specific cases.  

Taking the stellar wind to be isothermal, $u_{slow}$ can be calculated from a Parker wind~\citep{Parker1958}:
\begin{equation}
    \biggr(\frac{u}{c_s}\biggr)^2-ln\biggr(\frac{u}{c_s}\biggr)^2=4\biggr(\frac{r}{r_{cs}}\biggr)+4\biggr(\frac{r_{cs}}{r}\biggr)-3.
\end{equation}
At distances beyond the sonic point ($r_{cs}=G M_{\star}/2c_s$), where the wind becomes supersonic ($u>>c_s$), the approximation can be used~\citep{Blackman}:
\begin{equation}
    u_{slow} \approx 2 c_s \sqrt{\ln{(r/r_{cs})}},
    \label{eqn:parkerwind}
\end{equation}
where $c_s$ and $r_{cs}$ are the sonic speed and sonic radius, respectively. These depend on temperature, which can be related to rotation rate as a power law:
\begin{equation}
    T = T_{\odot} \biggr(\frac{\Omega}{\Omega_{\odot}}\biggr)^n,
    \label{eqn:temprelation}
\end{equation}
\citep{Ivanova2003,Holzwarth2007,See2014,Ahuir2020}. The wind speed is therefore also dependent on rotation rate and thus, stellar age. In this paper, we select $n = 0.6$ as a representative value. The variation of this power law was considered in more depth in Paper 1.
\\
Assuming a constant $\Delta \phi$, the variation in CIR radius with stellar age can then be modelled, as shown in Figures~\ref{fig:cirwithage}(b) and (c). Results are shown as a function of stellar radius (Figure~\ref{fig:cirwithage}(b)) and units of AU (Figure~\ref{fig:cirwithage}(c)). The minimum CIR radius is dominated by the $1/\Omega$ term in Equation~\ref{eqn:rcir}, rather than the wind velocity. This is consistent with results from previous work~\citep{Waugh2025}. For the youngest stars the minimum CIR radius is constant and close to the star, before the radius shrinks during spin up and has its smallest value at about 10 stellar radii or 0.05AU. During spin down, the minimum CIR radius moves out rapidly and continues to increase, albeit at a slower rate, once entering Skumanich spin down. The orbits of Mercury and Venus are shown on the plots for reference, by the brown and green points respectively. 

\section{How does the distribution of particles produced in shocks vary with stellar age?}

\begin{figure*}
    \centering \includegraphics[width=2.1\columnwidth]{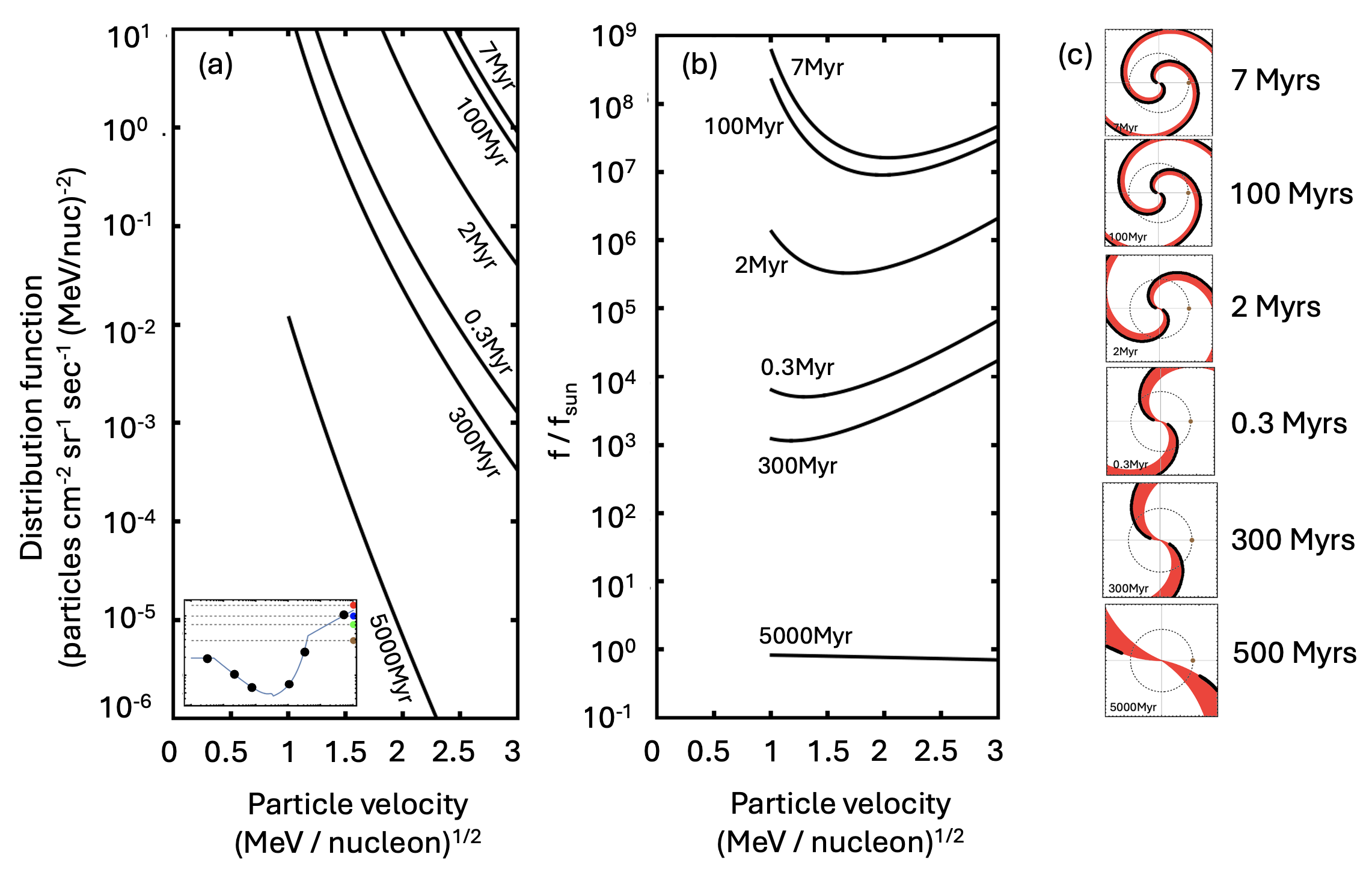}
    \caption{(a) The distribution of particle energies from the CIRs for various stellar ages. (b) As in (a) but as a fraction for the present day Sun. (c) The CIRs (black) and fast wind streams (red) for the selected ages. Mercury's orbit is shown for scale by the dashed circle.}
\label{fig:distfunctions}
\end{figure*}

\cite{FiskLee} published a model of the acceleration of particles within the shock of CIRs. Their model provides the distribution of particles at various speeds or energies, that would be detected at a planet of orbital radius, r. This is an approximate solution and is given by the equation below:
\begin{equation}
    f=\biggr(\frac{r}{r_{sh}}\biggr)^{2\beta/(1-\beta) + V/(\kappa_0 v)} v^{-3/(1-\beta)} \exp\biggr(-\frac{6\kappa_0\beta v}{V(1-\beta)^2}\biggr).
    \label{eqn:fisklee}
\end{equation}
where $r_{sh}$ is the radius of the shock, $V$ is the wind velocity and $\kappa_0$ is a constant with value $3.5\times10^5$. $\beta$ is a constant that relates to the ratio of magnetic field strength outside and inside the shock, we take it to be 1/3 as is consistent with \cite{FiskLee} and $v$ is the velocity of the shocked particle.\\

\begin{figure*}
    \centering \includegraphics[width=2\columnwidth]{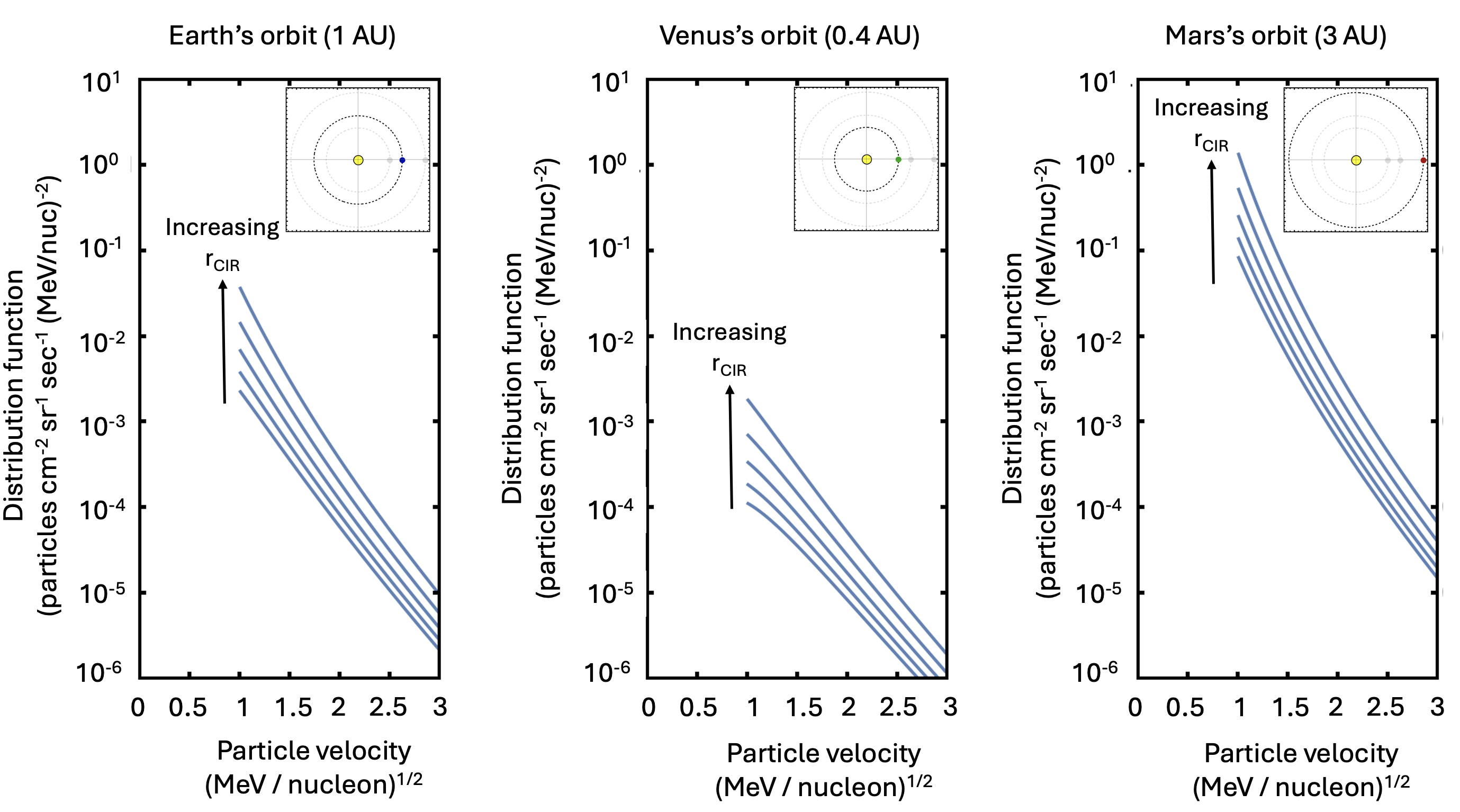}
    \caption{Distributions of particles produced in shocks for CIRs of different radii (1.5, 2, 2.5, 3, 3.5 AU), as experienced by planets in an Earth, Venus and Mars orbit.}
\label{fig:distfunction1}
\end{figure*}

Equation~\ref{eqn:fisklee} gives the distribution of particle energies from CIRs calculated for a planet in a 1 AU orbit, and we plot this in Figure~\ref{fig:distfunction1}(a) for various stellar ages. The inset panel of the figure shows the ages selected, on the plot of CIR radius with age. Note that at early ages (when the CIR radius is decreasing with age) curves are found at progressively higher values as time increases. At later times, however, when the CIR radius is increasing with age, the curves move to progressively lower values. In Figure~\ref{fig:distfunction1}(b) we show the distribution function as a fraction of the current day Sun (as calculated by this model). Whilst in (c) we show the shapes of the CIRs at the selected ages, with the CIRs in black and the fast wind stream in red. Within these plots we also show the orbit of Mercury (shown by the dashed circle and brown point) for scale. We discuss this Figure in more depth below.\\

The plot of minimum CIR radius with stellar age (shown in Figure~\ref{fig:cirwithage}) can be sampled at a few ages. This is shown in the inset of Figure~\ref{fig:distfunctions} (a) for the ages 0.3, 2, 7, 100, 300 and 5000 Myrs. The distribution of particle energies is then shown in the main plot of Figure~\ref{fig:distfunctions} (a) for these ages. Here we have used Equation \ref{eqn:fisklee}, varying the minimum radius of CIRs and the wind velocity with stellar age. The wind velocity in this model is varied using the Parker wind approximation (Equation~\ref{eqn:parkerwind}) and the temperature-rotation rate relation (Equation~\ref{eqn:temprelation}).
Equation \ref{eqn:fisklee} given by \cite{FiskLee} is an approximation to their model, which they note to be valid for velocities $v > 1$ (MeV/nucleon)$^{1/2}$, thus the curves here have been terminated at this point.\\


As the star initially ages, going from disc-locked to the spin-up and contraction stage, the distribution of particle energies shifts to the right, to higher velocities. Hence, at any particle velocity (e.g. 3 MeV) the number of particles increases. Thus, as the star spins up and the CIRs are brought closer in to the star, the planet experiences a greater number of higher energy particles from the CIR shocks. The star then begins saturated spin down and the minimum radius of CIRs increases sharply, this leads to the distribution of particle energies shifting left (to lower velocities) and the planet experiencing a drop in the number of high energy particles. Finally, the star enters spin down governed by the Skumanich relation and the minimum radius of CIRs continues to increase but more steadily. Again, this results in a drop in the numbers of high energy particles experienced by the planet.\\
Fig. \ref{fig:distfunctions}(a) shows that, from 0.3 Myrs to 2 Myrs, the number of particles with velocities of (for example) 3 (MeV/nucleon)$^{1/2}$ increases by a factor of 30 and then by a factor of 20 from 2 to 7 Myrs. Between 7 and 100 Myrs, the number of these particles drops only by a factor of 1.6, but in the next 200 Myrs it drops by a factor of 1700. As the star ages from 300 Myrs to 5000 Myrs (around the age of the current day Sun), the model suggests a drop in the 3 (MeV/nucleon)$^{1/2}$ by a factor of 25000, or 4 orders of magnitude. Thus, a planet could experience a large variation in the distribution of shocked particles from CIRs over its star's lifetime.\\

In Figure~\ref{fig:distfunctions} (b) we show the distribution of particle energies as a fraction of the value that our model generates for the present day Sun. As the Sun is 4490 Myrs old, it lies very close to the final age selected from Figure~\ref{fig:distfunctions} (a) (5000 Myr) and thus this line in Figure~\ref{fig:distfunctions} (b) is approximately unity. At the youngest age, the distribution of highest energy particles is 10$^5$ times that of the Sun, and by 7 Myrs it is over 10$^7$ times that of the Sun before it decreases again with age. The relative shape of these curves also changes with age. At 0.3 and 300 Myrs, when the star spins relatively slowly (albeit faster than the present day Sun), a larger percentage of the particles generated are higher energy particles than is the case for our Sun. At the ages 7 or 100 Myrs when the star is rotating most rapidly, the distribution of particles generated becomes biased towards lower energy particles, although the planet still experiences more than 10$^7$ times the number of 3 (MeV/nucleon)$^{1/2}$ particles than the Earth with the present day Sun.\\

The CIR locations are shown in Figure~\ref{fig:distfunctions} (c) for the various selected ages. The CIRs are shown by the thick black curves and the fast wind streams are shown in red. For scale, the orbit of Mercury is shown by the circle and brown point on each panel. At the youngest and oldest ages, the star rotates slowly and thus the CIRs and wind streams form shallow spirals and the wind streams are wide. At ages in between, the star rotates quickly and thus the CIRs and wind streams generate tight, thin spirals. The minimum CIR radius can also be clearly seen here to decrease with age before increasing again after the fourth panel and by 5000 Myrs the minimum CIR radius lies beyond the orbit of Mercury.

\section{What do planets experience as the star ages?}

The orbital distance of a planet also has consequences for the distribution of shocked particles that it experiences. As examples, we select orbital radii suitable for Venus, the Earth and Mars. We use the relation from \cite{FiskLee}. Following their approach, we select; $\beta= 1/3$; $\kappa_0 = 3.5\times10^5$; $V = 800$ km/s; and $r_{CIR}=(1.5, 2, 2.5, 3, 3.5)$ AU, we plot in Figure~\ref{fig:distfunction1} the distributions of particle energies from CIRS at a range of radii. Again, curves are truncated at $v =1$ (MeV/nucleon)$^{1/2}$. As the CIR moves out in radius, the distribution of particle energies shifts downwards, as fewer shocked particles reach the planet. It is notable that a relatively small increase in CIR radius (1.5 to 3.5 AU) can result in an order of magnitude difference in the distribution of particle energies. Planets at larger orbital radii experience a greater number of higher energy particles than those closer in.\\

\begin{figure}
    \centering \includegraphics[width=1\columnwidth]{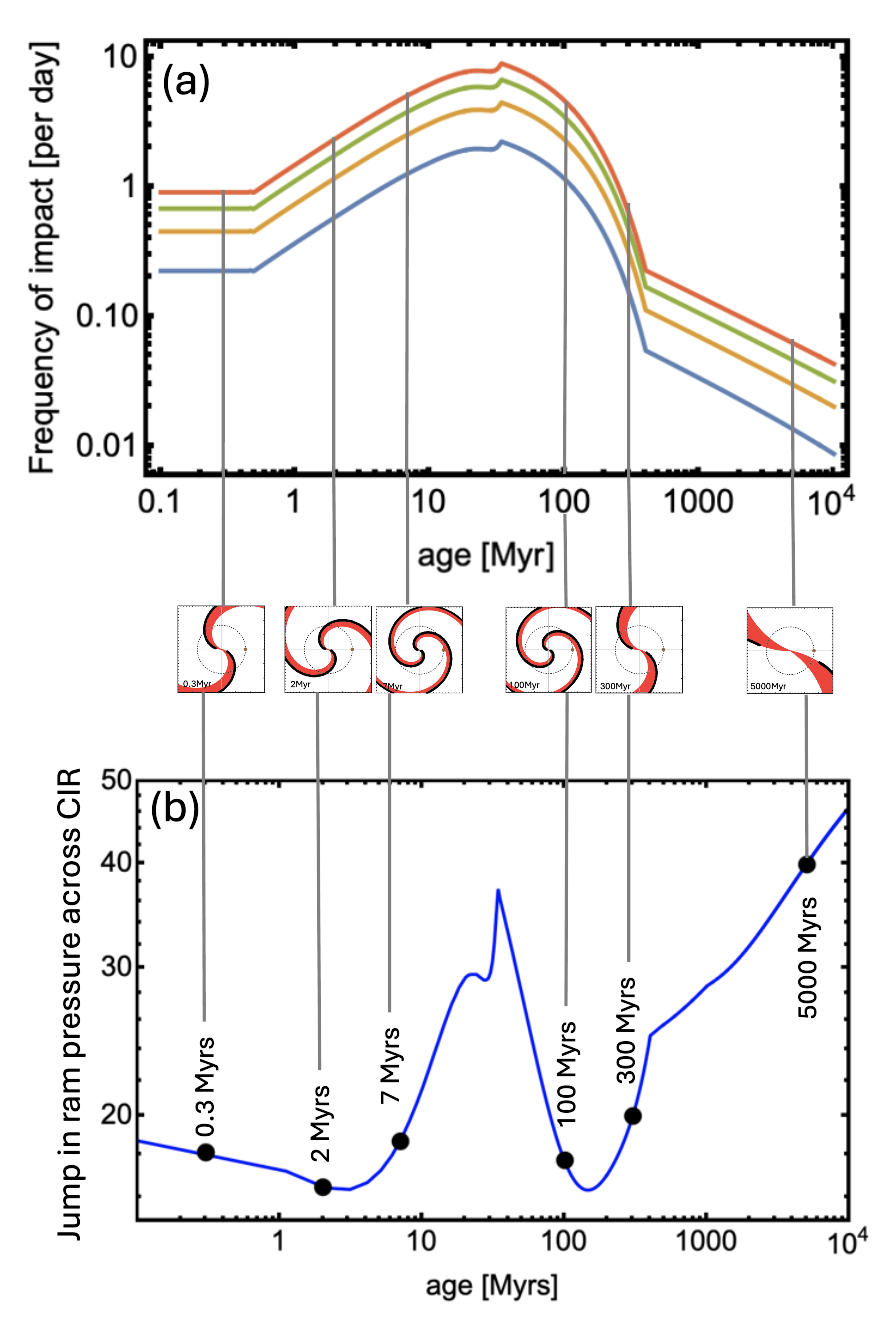}
    \caption{(a) Curves showing the frequency of CIR impacts on a planet in a 1 AU orbit, for 2 (blue), 4 (orange), 6 (green) and 8 (red) CIR streams. (b) The jump in ram pressure ($p_{ram,after}/p_{ram,before}$) of the CIR with stellar age. The CIR shapes with age are also shown for completeness.}
\label{fig:earththwacking}
\end{figure}

Figure~\ref{fig:earththwacking} shows the frequency with which a planet at 1 AU could experience CIR-collisions, assuming that the star consistently supports the CIRs throughout its life. We have assumed that the orbital radius of the planet does not change. The method for calculating CIR-planet interactions was shown in Paper 1~\citep{Waugh2025}, and is calculated as the beat frequency,
\begin{equation}
    f_{beat} = f_{CIR} - f_{planet},
\end{equation}
where $f_{CIR}$ and $f_{planet}$ are the frequencies of the CIR and planet, respectively. Here we use this to show the variation in CIR-planet collisions with stellar age. A greater number of CIR streams naturally results in a greater frequency of impact, and in Figure~\ref{fig:earththwacking}(a) the frequency of impact is shown for a star supporting two (blue curve), four (orange curve), six (green curve) and eight (red curve) CIR streams at a given time. The trend of the frequency of impact follows the shape of the rotation rate, showing an increase in frequency of collisions with age until the star begins to spin down. Increasing the number of streams increases the impact frequency by less than a factor of 10, and increases the frequency by the same factor at all ages. Whilst stars are unlikely to host 8 CIR streams at a given time, this figure shows the extent of diminishing return in which increasing the number of streams makes progressively less impact on the CIR-planet collision frequency. The ages sampled earlier in the paper (0.3, 2, 7, 100, 300 and 5000 Myrs) in Figure~\ref{fig:distfunctions} are shown again here by the black dashed lines. The CIR shapes are also shown again alongside for completeness. The correlation between CIR shape (or spiral tightness) and frequency of CIR impact can be easily seen. The CIR shape is tied to rotation rate through the Parker spiral expression:
\begin{equation}
   \phi = \phi_0 - \frac{\Omega}{u_{ft}} (r - 1) R_{\star},
\end{equation}
where $\phi$ and $r$ represent the longitudinal and radial coordinates, respectively, the streamline joins to the stellar surface at longitude $\phi_0$, and $\Omega$ is the stellar rotation rate. $u_{ft}$ is the fast wind velocity and the unit of $r$ here is stellar radii ($R_{\star}$).\\
For slow rotation rates the spiral is shallow, and this occurs at both young ages (before spin up) and late ages (in later stages of spin down). The star is rotating relatively slowly, and these shallow spirals that corotate with the star sweep past the planet relatively infrequently, leading to low frequencies of impact. During spin up of the star and early stages of spin down, the stellar rotation rate is high and the spiral is tight. The orbital period of the planet is constant, and thus so is $f_{planet}$, but $f_{CIR}$ evolves as the stellar frequency multiplied by the number of CIR streams ($N f_{\star}$). This increased stellar rotation rate leads to higher frequencies of CIR-planet collisions at these ages, and the CIRs are supported in tighter spirals. Whilst the CIR impacts are less frequent at very young and old ages, the changing tightness of the spirals means that the fast wind stream encompassing the CIR that is experienced by the planet is larger for these slow rotation rates than faster ones. This means it takes longer for the CIR to sweep past the planet.\\

The jump in ram pressure across the CIR can be calculated from the following relation:
\begin{equation}
\frac{p_{\rm {ram,after}}}{p_{\rm {ram,before}}} = 
    \frac{4(u_f/u_s - 1)^2}{(\gamma + 1)((\gamma-1) + 2/M_1^2)}
\end{equation}
where $\gamma$ is the ratio of specific heats (a value of 5/3 is used here) and $M_1$ is the upstream Mach number given by $M_1 = (u_{fast} - u_{slow})/c_{slow}$, where $c_{slow}$ is the sound speed in the slow wind. This was discussed in detail in Paper 1~\citep{Waugh2025}.

This jump in pressure is shown in Figure~\ref{fig:earththwacking}(b), again with the previously used ages shown by the black points. More discussion is given in Paper 1 as to the impact of various parameters on the pressure jump and here we show only one example with the following parameters, in order to show the trends with stellar age; $M_{\star} = 1 M_{\odot}$, $R_{\star} = 1 R_{\odot}$. The fast and slow wind temperatures are calculated from the temperature-rotation rate relation (Equation~\ref{eqn:temprelation}), scaled to the fast ($T_{\odot f}$) and slow ($T_{\odot s}$) wind temperatures of the solar wind:
$T_{slow} = T_{\odot s}\bigr(\Omega/\Omega_{\odot}\bigr)^{0.6}$,
$T_{fast} = T_{\odot f}\bigr(\Omega/\Omega_{\odot}\bigr)^{0.6}$, $T_{\odot f}=8\times10^5$ K, $T_{\odot s} =1\times10^5$ K. The wind temperatures, and therefore wind velocities, can then be cast in terms of rotation rate and thus the ram pressure jump $p_{ram,before}/p_{ram,after}$ can be calculated.\\
The ram pressure jump follows the same trends as the rotation rate and frequency of CIR-planet impact curves, but is almost constant over the ages considered here. Stellar ages where CIR-planet collisions are high are typically also ages at which the ram pressure jump is high. This means that at around a few tens of megayears, the planet experiences frequent and strong CIR impacts by thin wind streams. At early ages, the planet experiences less frequent and weaker CIR impacts by wider wind streams. At late ages, the planet experiences less frequent but strong CIR impacts, again by wide wind streams.

\section{For low mass stars, shocks form beyond the habitable zone}

\begin{figure}
    \centering \includegraphics[width=1\columnwidth]{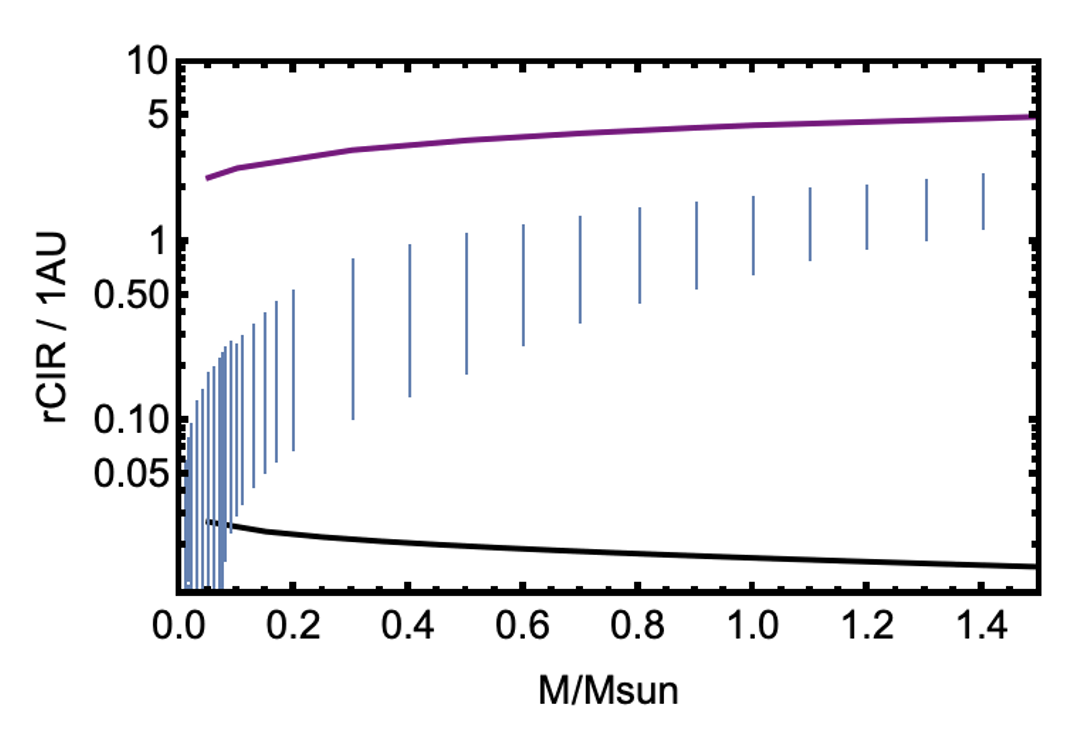}
    \caption{Minimum CIR radius with stellar mass. Blue lines show the locations of planetary habitability for various stars using data from \citet{2015A&A...577A..42B}, the black line shows the minimum CIR radius as calculated from Equation~\ref{eqn:rcir} for a star with $P_{\star}=3$ days, and the purple line shows the radius at which the CIR shock would form (where $M_1=1$) assuming 
  .}
\label{fig:stellarmass}
\end{figure}

Whilst above we have focused on varying the age of a solar mass star, the minimum radius at which CIRs form (Equation~\ref{eqn:rcir}) is dependent on the stellar mass through the sonic radius in the wind speed. This minimum CIR radius is shown by the black curve in Figure~\ref{fig:stellarmass}, for a star with a rotation period $P_{\star}=3$ days. The formation of a CIR does not necessarily mean the formation of a shock. Shocks will form within the CIR when the upstream Mach number is equal or greater than 1 ($M_1\geq1$). The radius at which the CIRs form shocks is also dependent on the stellar mass, again through the wind speeds. An example of this is shown in Figure~\ref{fig:stellarmass} by the purple curve, where the following parameters have been used; $T_{slow}=2\times10^5$ K and $T_{fast}=5\times10^5$ K. The shock radius also depends on the stellar radius, and thus we have used the mass-radius relation of $R = (M/M_{\odot})^{0.8} R_{\odot}$. The blue lines show the locations of planetary habitability, here defined as the planetary equilibrium temperature being consistent with supporting liquid water. For a planet radiating as a blackbody at temperature $T_p$ and with bond albedo $A$ then
\begin{equation}
    4\pi R_p^2\sigma T_p^4 = \frac{L_{\star}}{4\pi a^2}\pi R_p^2(1-A),
\end{equation}
where $R_p$ is the radius of the planet and $a$ is the orbital radius of the planet. Assuming that the exoplanet has the same equilibrium temperature as the Earth, and the exoplanet has the same bond albedo as the Earth (i.e. has the same greenhouse effect), the location of the habitable zone in units of AU is equal to the square root of the fractional luminosity:
\begin{equation}
\frac{a}{a_{\oplus}}=\biggr(\frac{L_{\star}}{L_{\odot}}\biggr)^{1/2} .
\end{equation}
Using evolutionary tracks and internal structure for brown dwarfs and low mass stars \citep{2015A&A...577A..42B}, the luminosity can then be related to the stellar mass and the blue lines plotted.\\

For all but the very lowest masses (less than about 0.1$M_{\odot}$), CIRs can form within the habitable zone. However, at all stellar masses the CIR shocks form \textit{beyond} the habitable zone. This is the regime seen within our Solar System, meaning that models such as~\cite{FiskLee} should be applicable to other star systems. This suggests that, as in our Solar System, the CIR may form within the orbit of the habitable zone but the shocks form further out and cause the energetic particles rain back inwards towards the inner stellar system.

\section{Conclusions}

This paper has focused on the evolution of CIRs over the lifetime of a solar mass star. The stellar evolution was modelled using a rotational evolution curve constructed from four functions that describe the phases of: disc locking, stellar contraction and spin up, saturated spin down and Skumanich spin down. The CIR locations were modelled using the method discussed in Paper 1~\citep{Waugh2025} and the distribution of high energy particles generated from the CIR was modelled using the method from~\cite{FiskLee}. We used a simple rotational evolution model for its transparency and reproducibility, the trends shown here motivate further study of particle distributions of these stars at young ages, when planetary atmospheres were forming. The results of this work are listed below.\\

\begin{itemize}
    \item The minimum radius at which CIRs form ($r_{CIR}$) is inversely related to the rotation rate, decreasing during spin up and increasing during spin down. The CIRs form closest to the star at young ages, after stellar contraction. For the fast rotator modelled here, CIRs form within 1 AU for ages less than about 5000Myrs.\\
    \item The particle distribution functions generated from the~\cite{FiskLee} model show large variations with stellar age. At the age of the current day Sun (roughly represented by the 5000 Myr curve in Figure~\ref{fig:distfunction1}) the CIR shocks do not produce a large percentage of high energy particles. However, the shape of this distribution changes with stellar age. At 0.3, 2 and 300Myrs the distribution produces a larger percentage of high energy particles than the present day Sun. Not only does it produce a larger percentage of high energy particles but they also generate a greater \textit{absolute} number. At 7 and 100Myrs, the CIRs also generate a greater absolute number of high energy particles than the present day Sun, however, the shape of the distribution then biases the lowest energy particles. Despite this, due to the extremely rapid rotation rates, the number of 3 (MeV/nucleon)$^{0.5}$ particles produced at these ages is around $10^7$ times that of the present day Sun. These ages also correspond to when the Parker spiral is tightest, meaning that the high energy particles raining back inwards into the inner Solar System and towards planets will occur for a longer percentage of the orbit.\\
    \item The frequency of CIR-planet collisions varies as the stellar rotation rate. At the earliest ages, the planet is hit relatively infrequently (every few days) by a weak pressure pulse in a wide, fast, wind stream. As the star spins up, the CIR impact rate increases, such that the planet is hit multiple times a day by a stronger pressure pulse in a thin wind stream. After spin down has begun and the rotation rate drops again, so too does the frequency of impact (down to once every hundred days at the latest ages) but by a stronger, wider CIR.\\
    \item For stars with masses of up to 1.4$M_{\odot}$, the shocks produced by CIRs form beyond the habitable zone whilst the CIRs themselves can begin to form well below this. This is consistent with what is observed within our Solar System, and thus solar models such as those by~\cite{FiskLee} should be applicable to other stellar systems too.
\end{itemize}


\section*{Acknowledgements}

The authors acknowledge support from STFC consolidated grant number ST/R000824/1.

\section*{Data Availability}
The research data from this paper can be accessed at: \url{https://doi.org/10.17630/4d379293-77d6-47d5-987e-b16c3210481c}.
\\

For the purpose of open access, the authors have applied a Creative Commons Attribution (CC BY) license to any Author Accepted Manuscript version arising.



\bibliographystyle{mnras}
\bibliography{REF} 

\bsp	
\label{lastpage}
\end{document}